\documentclass[aps,prl,reprint,groupedaddress]{revtex4-1}
\usepackage{amssymb}
\usepackage{amsmath}
\usepackage{array}
\usepackage{multirow,bigdelim}
\usepackage{enumitem}
\usepackage{xcolor}

\input{tcilatex}

\begin{document}

\title{A relation among tangle, 3-tangle, and von Neumann entropy of
entanglement for three qubits }
\author{ Dafa Li$^{1*}$, Maggie Cheng$^{2}$, Xiangrong Li $^{3}$, Shuwang Li$%
^{2}$}

\begin{abstract}
In this paper, we derive a general formula of the tangle for pure states of
three qubits, and present three explicit local unitary (LU) polynomial
invariants. Our result goes beyond the classical work of tangle, 3-tangle
and von Neumann entropy of entanglement for Ac\'{\i}n et al.' Schmidt
decomposition (ASD) of three qubits by connecting the tangle, 3-tangle, and
von Neumann entropy for ASD with Ac\'{\i}n et al.'s LU invariants. In
particular, our result reveals a general relation among tangle, 3-tangle,
and von Neumann entropy, together with a relation among their averages. The
relations can help us find the entangled states satisfying distinct
requirements for tangle, 3-tangle, and von Neumann entropy. Moreover, we
obtain all the states of three qubits of which tangles, concurrence,
3-tangle and von Neumann entropy don't vanish and these states are endurable
when one of three qubits is traced out. We indicate that for the three-qubit
W state, its average von Neumann entropy is maximal only within the W SLOCC
class, and that under ASD the three-qubit GHZ state is the unique state of
which the reduced density operator obtained by tracing any two qubits has
the maximal von Neumann entropy.
\end{abstract}

\maketitle

\affiliation{
$^1$Department of Mathematical Sciences, Tsinghua University,
Beijing, 100084, China. Corresponding author: lidafa@tsinghua.edu.cn\\
$^2$ Department of Applied Mathematics, Illinois Institute of Technology, Chicago, IL 60616, USA\\
$^3$ Department of Mathematics, University of California-Irvine, Irvine, CA 92697, USA }


\section{Introduction}

Quantum entanglement is considered as a unique quantum mechanical resource 
\cite{Nielsen}. Entanglement takes an important role in quantum information
and computation. Examples include quantum teleportation, quantum
cryptography, quantum metrology, and quantum key distribution. Considerable
efforts have been made to explore the entanglement classification via local
unitary operators (LU), local operations and classical communication (LOCC),
and Stochastic LOCC (SLOCC) \cite{Bennett}-\cite{Kraus-PRA}. It is known
that any two states of the same LU class have the same amount of
entanglement \cite{Dur, Verstraete, Kraus-prl}. Under SLOCC, pure states of
three qubits were partitioned into six equivalence classes: GHZ, W, A-BC,
B-AC, C-AB, and A-B-C \cite{Dur}. It has been established that two bipartite
states are LU equivalent if and only if their Schmidt coefficients coincide 
\cite{Nielsen, Kraus-PRA}. Ac\'{\i}n et al. proposed the Schmidt
decomposition for three qubits \cite{Acin00, Acin01}. Kraus introduced a
standard form for multipartite systems and showed that two states are LU
equivalent if and only if their standard forms coincide \cite{Kraus-prl,
Kraus-PRA}.

\vspace{5pt}

Coffman et al. defined the tangle for the reduced density operator $\rho
_{AB}$ (i.e., $tr_{C}\rho _{ABC}$)\ of a three-qubit state below \cite%
{Coffman}. Let 
\begin{equation}
\overline{\rho _{AB}}=\sigma _{y}\otimes \sigma _{y}\rho _{AB}^{\ast }\sigma
_{y}\otimes \sigma _{y},  \label{coffm}
\end{equation}%
where $\rho _{AB}^{\ast }$ is the complex conjugate of $\rho _{AB}$ and $%
\sigma _{y}$ is the Pauli matrix. Note that $\rho _{AB}\overline{\rho _{AB}}$
has only real and non-negative eigenvalues $\eta _{1}^{2}$, $\eta _{2}^{2}$, 
$\eta _{3}^{2}$, and $\eta _{4}^{2}$, where $\eta _{1}\geq \eta _{2}\geq
\eta _{3}\geq \eta _{4}$. Then, the entanglement tangle of $\rho _{AB}$ is
defined as 
\begin{equation}
\tau _{AB}=[\max \{\eta _{1}-\eta _{2}-\eta _{3}-\eta _{4},0\}]^{2}.
\label{tangle-1}
\end{equation}%
The tangle $\tau _{AC}$ of $\rho _{AC}$ ($=tr_{C}\rho _{ABC}$) and the
tangle $\tau _{BC}$ of $\rho _{BC}$ ($=tr_{A}\rho _{ABC}$) can be similarly
defined.

\vspace{5pt}The idea of taking average measure of entanglements comes from
an extremely useful technique in quantum information theory, the
\textquotedblleft average subsystem approach\textquotedblright\ proposed by
Page \cite{Page}. 
For a pure state $|\psi \rangle $ of three qubits, D\"{u}r et al. defined
its average residual entanglement as $\bar{\epsilon}(\psi )=\frac{1}{3}%
(\epsilon (\rho _{AB})+\epsilon (\rho _{AC})+\epsilon (\rho _{BC}))$, where $%
\epsilon (\rho _{xy})$ is some entanglement measure, where notation $xy$
means $AB,AC,$ or $BC$ \cite{Dur}. If $\epsilon (\rho _{xy})$ is the tangle
(the von Neumann entropy), then the average residual entanglement $\ \bar{%
\epsilon}(\psi )$ is the average tangle (von Neumann entropy). A state of
four qubits is defined to be maximally entangled if its average bipartite
entanglement (for example, the average tangle or the average entropy) with
respect to all possible bi-partite cuts is maximal \cite{Gour}. Significant
efforts have contributed to the average tangle \cite{Dur, Gour,Chen, Ana}
and von Neumann entropy, specially, the average entropy in bipartite,
tripartite, and multi-partite scenarios \cite%
{Soo,Kallin,Kumar,You,Koscik,Ana,Gour}, and reference \cite{PRL}. D\"{u}r et
al. showed that the W state of three qubits has the maximal average tangles
and indicated that for the GHZ state of three qubits, the tangle vanishes 
\cite{Dur}.

Considerable efforts have also been undertaken on the study of polynomial
invariants for $n$ qubits \cite{Linden, Sudbery, Leifer, Fei, Acin00} \cite%
{Poly}. SLOCC (LU) polynomial invariants can be used for SLOCC\ (LU)
entanglement classification of pure states of n qubits and can also be used
as entanglement measure.

\vspace{5pt}

In this paper, we derive a general formula of the tangle for three qubits,
and obtain three LU polynomial invariants of degree 4. We calculate tangle,
average tangle, 3-tangle, von Neumann entropy , and average von Neumann
entropy in the framework of ASD. We derive an equation which tangle,
3-tangle, and von Neumann entropy satisfy. Via the equation, we present
relations among tangles and von Neumann entropy, among tangle, 3-tangle, and
von Neumann entropy , among the average tangles, the average von Neumann
entropy , and 3-tangle, and among von Neumann entropy and Ac\'{\i}n et al.'s
LU invariants. The relations can help us find the entangled states
satisfying distinct requirements for tangle, 3-tangle, and von Neumann
entropy.

We obtain all three-qubit states whose tangle, concurrence, 3-tangle and von
Neumann entropy do not vanish. Via von Neumann entropy for ASD, we indicate
that the GHZ state is the unique three-qubit state under ASD that has the
maximal von Neumann entropy of $\ln 2$, and the average von Neumann entropy
of the W state is maximal only within the W SLOCC\ class.

\section{Tangles for ASD}

In this section, we derive the formulas for tangles for pure states of three
qubits and for the ASD, and present a relation between tangles and Ac\'{\i}n
et al.'s LU polynomial invariants.

\subsection{Homogeneous polynomial of degree 4 for tangle}

Let $|\psi \rangle =\sum_{i=0}^{7}c_{i}|i\rangle $. By solving Eq. (\ref%
{tangle-1}), we obtain $\tau _{AB}$, $\tau _{AC}$, and $\tau _{BC}$ as
follows, 
\begin{eqnarray}
\tau _{AB} &=&\Delta -\frac{\tau _{ABC}}{2},  \label{tangle-ab} \\
\tau _{AC} &=&\Phi -\frac{\tau _{ABC}}{2},  \label{tangle-ac} \\
\tau _{BC} &=&\Psi -\frac{\tau _{ABC}}{2},  \label{tangle-bc}
\end{eqnarray}%
where $\Delta $, $\Phi $, $\Psi $ are defined in Eqs. (\ref{tang-ab}, \ref%
{tang-ac}, \ref{tang-bc})\ {\ in Appendix A. Following \cite{Dli-12}, the
3-tangle $\tau _{ABC}$ can be} written as 
\begin{eqnarray}
\tau _{ABC} &=&4|(c_{0}c_{7}-c_{2}c_{5}-c_{1}c_{6}+c_{3}c_{4})^{2}  \notag \\
&&-4(c_{0}c_{3}-c_{1}c_{2})(c_{4}c_{7}-c_{5}c_{6})|\text{.}
\label{tangle-abc}
\end{eqnarray}

Note that $\Delta $, $\Phi $, and $\Psi $ are LU homogeneous polynomial
invariants of degree 4 in the state coefficients and their complex
conjugates. Therefore, one needs only "$+,-,\times $" operations of the
state coefficients and their complex conjugates to compute the tangles.

Williamson et al. obtained $I(\tilde{_{a}}\tilde{_{b}})=tr[\rho _{ab}\tilde{%
\rho _{ab}}]=\tau _{ab}+\frac{1}{2}\tau _{abc}$, where $\tilde{\rho _{ab}}%
=\sigma _{y}\otimes \sigma _{y}\rho _{ab}^{T}\sigma _{y}\otimes \sigma _{y}$ 
{\cite{Williamson}}. Note that Coffman et al. used $\rho _{ab}^{\ast }$
(which is the complex conjugate of $\rho _{ab}$) in Eq. (\ref{coffm}) rather
than $\rho _{ab}^{T}$. It is known that it is complicated to compute the
reduced density operator $\rho _{ab}$ and $tr[\rho _{ab}\tilde{\rho _{ab}}]$%
. Comparatively, it is easier and simpler to compute $\Delta $ than $tr[\rho
_{ab}\tilde{\rho _{ab}}]$.

\subsection{Tangles for ASD}

Ac\'{\i}n et al. proposed the following Schmidt decomposition for pure
states of three qubits, 
\begin{eqnarray}
|\psi \rangle &=&\lambda _{0}|000\rangle +\lambda _{1}e^{i\phi }|100\rangle
+\lambda _{2}|101\rangle  \notag \\
&&+\lambda _{3}|110\rangle +\lambda _{4}|111\rangle ,  \label{sd-1}
\end{eqnarray}%
where $\lambda _{i}\geq 0$, $i=0,1,\cdots ,4$, $0\leq \phi <2\pi $, and $%
\sum \lambda _{i}^{2}=1$ \cite{Acin00, Acin01}. Equation (\ref{sd-1}) is
referred to the ASD of $|\psi \rangle $.{} It is known that any state of
three qubits is LU equivalent to its ASD. For ASD, 3-tangle $\tau _{ABC}$ in
Eq. ({\ref{tangle-abc}}) can be reduced to 
\begin{equation}
\tau _{ABC}=4\lambda _{0}^{2}\lambda _{4}^{2}.  \label{3-tangle}
\end{equation}%
{$\tau _{AB}$ in Eq. (\ref{tangle-ab}) and $\tau _{AC}$ in Eq. (\ref%
{tangle-ac}) can be reduced similarly, 
\begin{align}
\tau _{AB}& =4\lambda _{0}^{2}\lambda _{3}^{2},  \label{2-tangle-ab} \\
\tau _{AC}& =4\lambda _{0}^{2}\lambda _{2}^{2}.  \label{2-tangle-ac}
\end{align}%
} 

{To reduce $\tau _{BC}$ in Eq. (\ref{tangle-bc}) for ASD, we write $\Psi $
in Eq. (\ref{tang-bc}) as $\Psi =2\Pi $, where 
$\Pi =[\lambda _{0}^{2}\lambda _{4}^{2}+2(\lambda _{1}\lambda _{4}-\lambda
_{2}\lambda _{3})^{2}+8\lambda _{1}\lambda _{2}\lambda _{3}\lambda _{4}\sin
^{2}\frac{\phi }{2}]. $ 
Then, we have 
\begin{eqnarray}
\tau _{BC} &=&2\left( \Pi -\lambda _{0}^{2}\lambda _{4}^{2}\right)  \notag \\
&=&4[(\lambda _{1}\lambda _{4}-\lambda _{2}\lambda _{3})^{2}+4\lambda
_{1}\lambda _{2}\lambda _{3}\lambda _{4}\sin ^{2}\frac{\phi }{2}]  \notag
\label{BC-1} \\
&=&4|\lambda _{1}\lambda _{4}e^{i\phi }-\lambda _{2}\lambda _{3}|^{2}.
\label{2-tangle-bc}
\end{eqnarray}
}

In Table 1, we summarize the tangles $\tau _{AB}$, $\tau _{AC}$, $\tau _{BC}$%
, and 3-tangle $\tau _{ABC}$ for ASD. 
\begin{table}[h]
\caption{$\protect\tau _{AB}$, $\protect\tau _{AC}$, and $\protect\tau _{BC}$
for ASD}
\label{T1}\centering
\begin{tabular}{|l|l|l|}
\toprule &  &  \\ 
$\tau _{AB}=$ & $4\lambda _{0}^{2}\lambda _{3}^{2}=$ & $4J_{3}$ \\ 
&  &  \\ 
\colrule &  &  \\ 
$\tau _{AC}=$ & $4\lambda _{0}^{2}\lambda _{2}^{2}=$ & $4J_{2}$ \\ 
&  &  \\ 
\colrule &  &  \\ 
$\tau _{BC}=$ & $4|\lambda _{1}\lambda _{4}e^{i\phi }-\lambda
_{2}\lambda_{3}|^{2}=$ \hspace{3ex} & $4J_{1}$ \\ 
&  &  \\ 
\colrule &  &  \\ 
$\tau _{ABC}=$ & $4\lambda _{0}^{2}\lambda _{4}^{2}=$ & $4J_{4}$ \\ 
&  &  \\ 
\botrule &  & 
\end{tabular}%
\end{table}

From Table \ref{T1}, it is straightforward to calculate the tangles and
3-tangle for any ASD. Because of the LU equivalence between a state and its
ASD, a state and its ASD have the same tangles, 3-tangle. For example, the
ASD of the W state is $(1/\sqrt{3})(|000\rangle +|101\rangle +|110\rangle )$%
. From Table 1, we conclude that the tangles of the W state $\displaystyle%
\tau _{AB}=\tau _{AC}=\tau _{BC}=\frac{4}{9}$, and the 3-tangle $\tau _{ABC}$
vanishes.

For ASD, Ac\'{\i}n et al. proposed five LU invariants \ $J_{i},i=1,2,3,4,5$,
where $J_{1}=|\lambda _{1}\lambda _{4}e^{i\phi }-\lambda _{2}\lambda
_{3}|^{2}$, and $J_{i}=(\lambda _{0}\lambda _{i})^{2}$ for $i=2,3,4$ \cite%
{Acin00, Acin01}. Thus, for ASD, the tangles and 3-tangle in Eqs. (\ref%
{2-tangle-ab}, \ref{2-tangle-ac}, \ref{2-tangle-bc}, \ref{3-tangle}) can be
rewritten in terms of LU invariants $J_{i}$ (see Table \ref{T1}), 
\begin{eqnarray}
\tau _{AB} &=&4J_{3}, \\
\tau _{AC} &=&4J_{2}, \\
\tau _{BC} &=&4J_{1}, \\
\tau _{ABC} &=&4J_{4}.
\end{eqnarray}

\section{von Neumann entropy for ASD and a relation between tangles and von
Neumann entropy}

\subsection{von Neumann entropy for ASD}

von Neumann entropy is defined as 
\begin{equation}
S(\rho )=-\sum \eta _{i}\ln \eta _{i},  \label{von-1}
\end{equation}%
where $\eta _{i}\geq 0$ are the eigenvalues of $\rho $, and $\sum_{i}\eta
_{i}=1$. Note that $0\ln 0=0$. 
Via ASD, it is easy to verify that 
\begin{eqnarray}
S(\rho _{A}) &=&S(\rho _{BC}),  \label{rdo-a} \\
S(\rho _{B}) &=&S(\rho _{AC}),  \label{rdo-b} \\
S(\rho _{C}) &=&S(\rho _{AB}).  \label{rdo-c}
\end{eqnarray}

It is well known that the eigenvalues of $\rho $ is just the roots of the
characteristic polynomial of $\rho $ . A tedious calculation derives the
characteristic polynomials of the reduced density operators $\rho _{\mu }$, $%
\mu =A$, $B$, $C$, which are $X^{2}-X+\alpha _{\mu }$, where $\alpha _{\mu }$
is just the sum of tangles and 3-tangle, ref. Table II. Thus, we establish a
relation among the entanglement measures: the von Neumann entropy, tangle
and 3-tangle.

From results in Table \ref{T1} and via ASD, a calculation yields Table \ref%
{T2}, in which abbreviation CP stands for the characteristic polynomial of
the reduced density operator, $\rho _{\mu }$. 
\begin{table}[th]
\caption{$S(\protect\rho _{\protect\mu })$ for ASD}
\label{T2}\centering
\begin{tabular}{|l|l|l|ll|}
\toprule &  &  &  &  \\ 
$\mu$ & RDO & CP of $\rho _{\mu }$ & $\alpha _{\mu }$ &  \\ 
&  &  &  &  \\ 
\colrule &  &  &  &  \\ 
$A$ & $\rho _{A}$ & $X^{2}-X+\alpha _{A}$ & $\alpha _{A}$ & $=
J_{2}+J_{3}+J_{4}$ \\ 
&  &  &  & $=\frac{\tau _{AB}+\tau _{AC}+\tau _{ABC}}{4}$ \\ 
&  &  &  & $=\frac{\tau _{A(BC)}}{4}$ \\ 
&  &  &  &  \\ 
\colrule &  &  &  &  \\ 
$B$ & $\rho _{B}$ & $X^{2}-X+\alpha _{B}$ & $\alpha _{B}$ & $=
J_{1}+J_{3}+J_{4}$ \\ 
&  &  &  & $=\frac{\tau _{AB}+\tau _{BC}+\tau _{ABC}}{4}$ \\ 
&  &  &  & $=\frac{\tau _{B(AC)}}{4}$ \\ 
&  &  &  &  \\ 
\colrule &  &  &  &  \\ 
$C$ & $\rho _{C}$ & $X^{2}-X+\alpha _{C}$ & $\alpha _{C} $ & $=
J_{1}+J_{2}+J_{4}$ \\ 
&  &  &  & $=\frac{\tau _{AC}+\tau _{BC}+\tau _{ABC}}{4} $ \\ 
&  &  &  & $=\frac{\tau _{C(AB)}}{4}$ \\ 
&  &  &  &  \\ 
\botrule &  &  &  & 
\end{tabular}%
\end{table}

\vspace{10pt} In Table \ref{T2}, the notation $\tau _{A(BC)}$ stands for the
tangle between a qubit A and a qubit pair BC, where the qubit pair BC is
considered a single object \cite{Coffman}. \vspace{5pt}

Let $\eta _{\mu }^{(1)}$ and $\eta _{\mu }^{(2)}$ be the two eigenvalues of $%
\rho _{\mu }$. We have 
\begin{eqnarray}
\eta _{\mu}^{(1)} &=\frac{1+\sqrt{1-4\alpha _{\mu }}}{2}, \\
\eta _{\mu}^{(2)} &=\frac{1-\sqrt{1-4\alpha _{\mu }}}{2}.
\end{eqnarray}

Then, from Eq. (\ref{von-1}), one can see that 
\begin{eqnarray}
S(\rho _{\mu }) &=&-(\eta _{\mu }^{(1)}\ln \eta _{\mu }^{(1)}+\eta _{\mu
}^{(2)}\ln \eta _{\mu }^{(2)}),  \label{entropy-1}
\end{eqnarray}

where $\mu \in \{A,B,C\}$ and $0\leq \alpha _{\mu }\leq 1/4$. \vspace{5pt}


Equation (\ref{entropy-1}) and Table \ref{T2} reveal a relation among
tangles, 3-tangle and von Neumann entropy . In other words, tangles,
3-tangle, and von Neumann entropy for any pure three-qubit state must
satisfy Eq. (\ref{entropy-1}). See Table \ref{T5} for the GHZ state, the W
state, as well as other states. We will explore the relation for details
below.

For $\mu \in \{A,B,C\}$, the derivative of the von Neumann entropy $S(\rho
_{\mu })$ with respect to $\alpha _{\mu }$ is given by 
\begin{equation}
(S(\rho _{\mu }))_{\alpha _{\mu }}^{\prime }=-\frac{1}{\sqrt{1-4\alpha _{\mu
}}}\ln \frac{1-\sqrt{1-4\alpha _{\mu }}}{1+\sqrt{1-4\alpha _{\mu }}}
\label{deriv-1}
\end{equation}%
$S(\rho _{\mu }))_{\alpha _{\mu }}^{\prime }>0$ when $0<\alpha _{\mu }<1/4$.
Hence, $S(\rho _{\mu })$ increases monotonically as $\alpha _{\mu }$
increases. Recall that $0\leq \alpha _{\mu }\leq 1/4$. Thus, $0\leq S(\rho
_{\mu })\leq \ln 2$, and $S(\rho _{\mu })=\ln 2$ \textit{iff} $\alpha _{\mu
}=1/4$, and $S(\rho _{\mu })=0\ $\textit{iff} $\alpha _{\mu }=0$. This means
that the maximal von Neumann entropy for one particle $\rho _{\mu }$ is $\ln
2$. \vspace{5pt}

For the GHZ state, $\alpha _{\mu }=\frac{1}{4}$ and $S(\rho _{\mu })=\ln 2$;
and for the W state, $\alpha _{\mu }=\frac{2}{9}$ and $S(\rho _{\mu })=\frac{%
3\ln 3-2\ln 2}{3}$. Thus, for three qubits, the GHZ state has the maximal
von Neumann entropy for any kind of the reduced density operator $\rho _{\mu
}$. 

\vspace{5pt}

\subsection{ Relation between tangles and von Neumann entropy}

\textit{Proposition 1.} For any pure state of three qubits, two different
types of tangles are equal \textit{iff} the corresponding the von Neumann
entropy are equal. \ That is, 
\begin{eqnarray}
(1)\quad S(\rho _{AC}) &=&S(\rho _{BC})\quad \mathit{iff}\quad \tau
_{AC}=\tau _{BC},  \notag \\
(2)\quad S(\rho _{AB}) &=&S(\rho _{BC})\quad \mathit{iff}\quad \tau
_{AB}=\tau _{BC},  \notag \\
(3)\quad S(\rho _{AB}) &=&S(\rho _{AC})\quad \mathit{iff}\quad \tau
_{AB}=\tau _{AC}.  \notag
\end{eqnarray}%
Clearly, $S(\rho _{AB})=S(\rho _{AC})=S(\rho _{BC})$ \textit{iff} $\tau
_{AB}=\tau _{AC}=\tau _{BC}$ (see Table III). \vspace{5pt}

Let $uvw$ and $xyz$ be two different three-character strings from the set $%
\{ABC,BAC,CAB\}$. Then, also $\tau _{u(vw)}=\tau _{x(yz)}$ \textit{iff} $%
S(\rho _{u})=S(\rho _{x})$.

{\underline{Proof of (1)}}: If $\tau _{AC}=\tau _{BC}$, then from Table \ref%
{T2}, $\alpha _{A}=\alpha _{B}$, and then $S(\rho _{A})=S(\rho _{B})$, which
leads to $S(\rho _{BC})=S(\rho _{AC})$ from Eqs. (\ref{rdo-a} -- \ref{rdo-b}%
). Conversely, if $S(\rho _{AC})=S(\rho _{BC})$, then $S(\rho _{B})=S(\rho
_{A})$, then $\alpha _{B}=\alpha _{A}$ because $S(\rho )$ is strictly
increasing, and then $\tau _{AC}=\tau _{BC}$ from Table \ref{T2}. Similarly,
claims (2) and (3) also hold.

\vspace{5pt}

\textit{Proposition 2.} For any pure state of three qubits, two different
types of tangles satisfy $\tau _{uv}>\tau _{xy}$ \textit{iff} the
corresponding the von Neumann entropy satisfy $S(\rho _{uv})<S(\rho _{xy})$.
That is, 
\begin{eqnarray}
(1)\quad \tau _{AC} &>&\tau _{BC}\quad \mathit{iff}\quad S(\rho
_{AC})<S(\rho _{BC}),  \notag \\
(2)\quad \tau _{AB} &>&\tau _{BC}\quad \mathit{iff}\quad S(\rho
_{AB})<S(\rho _{BC}),  \notag \\
(3)\quad \tau _{AB} &>&\tau _{AC}\quad \mathit{iff}\quad S(\rho
_{AB})<S(\rho _{AC}).  \notag
\end{eqnarray}

Let $uvw$ and $xyz$ be two different three-character strings from the set $%
\{ABC,BAC,CAB\}$. Then, also $\tau _{u(vw)}>\tau _{x(yz)}$ \textit{iff} $%
S(\rho _{u})>S(\rho _{x})$.

{\underline {Proof of (1)}}: If $\tau _{AC}>\tau _{BC}$, then $\alpha
_{A}>\alpha _{B}$ from Table II, and then $S(\rho _{A})>S(\rho _{B})$, i.e $%
S(\rho _{BC})>S(\rho _{AC})$, because $S(\rho )$ is strictly increasing.
Conversely, if $S(\rho _{AC})<S(\rho _{BC})$, i.e. $S(\rho _{B})<S(\rho
_{A}) $, then $\alpha _{B}<\alpha _{A}$ because $S(\rho )$ is strictly
increasing, and then $\tau _{AC}>\tau _{BC}$. Similarly, claims (2) and (3)
also hold. \vspace{5pt}

Next, we study the relation between the difference of two tangles and the
difference of the corresponding von Neumann entropy s. Using the
differential Mean Value theorem, we have 
\begin{eqnarray}
S(\rho _{A})-S(\rho _{B}) &=&(S(\rho _{\mu }))_{\alpha _{\mu }}^{\prime
}(\xi _{1})(\alpha _{A}-\alpha _{B})  \label{mean-1} \\
S(\rho _{A})-S(\rho _{C}) &=&(S(\rho _{\mu }))_{\alpha _{\mu }}^{\prime
}(\xi _{2})(\alpha _{A}-\alpha _{C})  \label{mean-2} \\
S(\rho _{B})-S(\rho _{C}) &=&(S(\rho _{\mu }))_{\alpha _{\mu }}^{\prime
}(\xi _{3})(\alpha _{B}-\alpha _{C})  \label{mean-3}
\end{eqnarray}

From Eq. (\ref{deriv-1}), we have $(S(\rho _{\mu }))_{\alpha _{\mu
}}^{\prime }(\xi _{i})>0$, $i=1,2,3$. Clearly, we can also use Eqs. (\ref%
{mean-1}, \ref{mean-2}, \ref{mean-3}) to prove Propositions 1 and 2. We next
calculate $S(\rho _{\mu }))_{\alpha _{\mu }}^{\prime }$ using the second
order Taylor expansion of $\ln (1\pm x)$, 
\begin{eqnarray*}
&&(S(\rho _{\mu }))_{\alpha _{\mu }}^{\prime } \\
&=&-\frac{1}{\sqrt{1-4\alpha _{\mu }}}[\ln (1-\sqrt{1-4\alpha _{\mu }})-\ln
(1+\sqrt{1-4\alpha _{\mu }})] \\
&\approx &-\frac{1}{\sqrt{1-4\alpha _{\mu }}}\times \\
&&[(-\sqrt{1-4\alpha _{\mu }}-\frac{1-4\alpha _{\mu }}{2})-(\sqrt{1-4\alpha
_{\mu }}-\frac{1-4\alpha _{\mu }}{2})] \\
&=&2.
\end{eqnarray*}

Therefore, $(S(\rho _{\mu }))_{\alpha _{\mu }}^{\prime }(\xi _{i})\approx 2$%
. Via Eqs. (\ref{mean-1}, \ref{mean-2}, \ref{mean-3}) we arrive at the
following Proposition. \vspace{5pt}

\textit{Proposition 3.} For any pure state of three qubits, the difference
of two tangles is approximately twice as large as the difference of the
corresponding von Neumann entropy . We further explain this in details
below. \vspace{5pt}

Let different $uvw$ and $xyz$ belong to $\{ABC,BAC,CAB\}$. Then, 
\begin{equation*}
\tau _{u(vw)}-\tau _{x(yz)}=\tau _{yz}-\tau _{vw}\approx 2[S(\rho
_{vw})-S(\rho _{yz})].
\end{equation*}%
That is, 
\begin{eqnarray}
\tau _{A(BC)}-\tau _{B(AC)} &=&\tau _{AC}-\tau _{BC}\approx 2[S(\rho
_{BC})-S(\rho _{AC})],  \notag \\
\tau _{A(BC)}-\tau _{C(AB)} &=&\tau _{AB}-\tau _{BC}\approx 2[S(\rho
_{BC})-S(\rho _{AB})],  \notag \\
\tau _{B(AC)}-\tau _{C(AB)} &=&\tau _{AB}-\tau _{AC}\approx 2[S(\rho
_{AC})-S(\rho _{AB})].  \notag
\end{eqnarray}

\subsection{Relation between the von Neumann entropy and Ac\'{\i}n et al.'s
LU invariants}

It is well known that LU invariants are considered as entanglement measure 
\cite{Sudbery, Acin00, Acin01}. So far, no one discusses the relations
between LU invariants (as measures) and other entanglement measures. Here,
we establish the relation between von Neumann entropy and LU invariants (as
entanglement measure) $J_{i}$, $i=1,2,3,4$. From Eq. (\ref{entropy-1}) and
Table \ref{T2}, we can write the von Neumann entropy with Ac\'{\i}n et al.'s
LU\ invariants $J_{i}$, $i=1,2,3,4$. Thus, we have the following immediate
results.

(1). For any pure state of three qubits, two different types of Ac\'{\i}n et
al.'s LU invariants are equal \textit{iff} the corresponding the von Neumann
entropy are equal. \ That is, $S(\rho _{A})=S(\rho _{B})$ \textit{iff} $%
J_{2}=J_{1}$; $S(\rho _{A})=S(\rho _{C})$ iff $J_{3}=J_{1}$; $S(\rho
_{B})=S(\rho _{C})$ \textit{iff} $J_{3}=J_{2}$.

(2). $S(\rho _{A})>S(\rho _{B})$ \textit{iff} $J_{2}>J_{1}$; $S(\rho
_{A})>S(\rho _{C})$ iff $J_{3}>J_{1}$; $S(\rho _{B})>S(\rho _{C})$ \textit{%
iff} $J_{3}>J_{2}$.

Via Eqs. (\ref{mean-1}, \ref{mean-2}, \ref{mean-3}) and $(S(\rho _{\mu
}))_{\alpha _{\mu }}^{\prime }(\xi _{i})\approx 2$, then we obtain the
following.

(3). For any pure state of three qubits, the difference of two von Neumann
entropy is approximately twice as large as the difference of the
corresponding Ac\'{\i}n et al.'s LU invariants. 
\begin{eqnarray}
S(\rho _{A})-S(\rho _{B}) &\approx &2(J_{2}-J_{1}) \\
S(\rho _{A})-S(\rho _{C}) &\approx &2(J_{3}-J_{1}) \\
S(\rho _{B})-S(\rho _{C}) &\approx &2(J_{3}-J_{2})
\end{eqnarray}

\section{Relation among tangle, 3-tangle, and von Neumann entropy}

Equation (\ref{entropy-1}) can be reduced to 
\begin{eqnarray}
S(\rho _{\mu }) &=&-[\eta _{\mu }^{(1)}\ln (1+\sqrt{1-4\alpha _{\mu }}) 
\notag \\
&&+\eta _{\mu }^{(2)}\ln (1-\sqrt{1-4\alpha _{\mu }})-\ln 2].
\label{Taylor-}
\end{eqnarray}%
Then by the second order Taylor expansion of $\ln (1\pm x)$, we approximate $%
S(\rho _{\mu })$ as follows,

\begin{eqnarray}
S(\rho _{\mu }) &\approx &-[\eta _{\mu }^{(1)}(\sqrt{1-4\alpha _{\mu }}-%
\frac{1-4\alpha _{\mu }}{2})  \notag \\
&&+\eta _{\mu }^{(2)}(-\sqrt{1-4\alpha _{\mu }}-\frac{1-4\alpha _{\mu }}{2}%
)-\ln 2]  \notag \\
&=&\ln 2-\frac{1}{2}+2\alpha _{\mu },  \label{Taylor-1}
\end{eqnarray}%
where $\mu \in \{A,B,C\}$. From Eq. ( \ref{Taylor-1}), we obtain the
following relation among tangle, 3-tangle, and the von Neumann entropy : 
\begin{eqnarray*}
2S(\rho _{A}) &\approx &2\ln 2-1+\tau _{A(BC)} \\
&=&2\ln 2-1+\tau _{AB}+\tau _{AC}+\tau _{ABC}, \\
2S(\rho _{B}) &\approx &2\ln 2-1+\tau _{B(AC)} \\
&=&2\ln 2-1+\tau _{AB}+\tau _{BC}+\tau _{ABC}, \\
2S(\rho _{C}) &\approx &2\ln 2-1+\tau _{C(AB)} \\
&=&2\ln 2-1+\tau _{AC}+\tau _{BC}+\tau _{ABC}.
\end{eqnarray*}

The states in Table III satisfy the above relations. The above relations can
help us find the states which satisfy special requirements for tangle,
3-tangle, and von Neumann entropy, or the states of which tangle, 3-tangle,
and von Neumann entropy are as big as possible. For example, for the state $%
|\kappa \rangle $ in Table III, $S(\rho _{A})=0.687$, $\tau _{AB}=\tau
_{AC}=4/9$, and $\tau _{ABC}=8/81$.

Next, we demonstrate how to use the above relation to explore properties of
tangle, 3-tangle, and von Neumann entropy. Let $S(\rho _{A})=\ln 2$
(maximum). Then, tangles $\tau _{AB}$ and $\tau _{AC}$, and 3-tangle $\tau
_{ABC}$ must satisfy the following. 
\begin{equation}
\tau _{AB}+\tau _{AC}+\tau _{ABC}\approx 1.  \label{tan-3-tang}
\end{equation}

For example, if $\tau _{ABC}=1$, then $\tau _{AB}$ and $\tau _{AC}$ vanish;
if $\tau _{ABC}=1/2$, then $\tau _{AB}+\tau _{AC}=1/2$.

\section{Relation among the average tangle, the average von Neumann entropy,
and 3-tangle}

Lots of efforts have contributed to investigate the the average tangle \cite%
{Dur, Gour} and the average von Neumann entropy \cite{PRL}. So far, no one
discusses the relation among them. In this subsection, we establish the
relation.

\subsection{Definition for the average tangle}

Let $A(\psi)$ be the average tangles for the state $|\psi \rangle $. Then, 
\begin{equation}
A(\psi)=\frac{\tau _{AB}+\tau _{AC}+\tau _{BC}}{3}.  \label{average-1}
\end{equation}

We explain how to calculate the average tangle $A$. First, take partial
traces over qubit A (resp. B and C) to get\ the reduced density operators $%
\rho _{BC}$ (resp. $\rho _{AC}$, and $\rho _{AB}$), then by the definition
of the tangle in Eq. (\ref{tangle-1}) calculate the tangle $\tau _{AB}$
(resp. $\tau _{AC}$, and $\tau _{BC}$) for the reduced density operator $%
\rho _{AB}$ (resp. $\rho _{AC}$, and $\rho _{BC}$). Finally make the average
of the tangles $\tau _{AB}$, $\tau _{AC}$, and $\tau _{BC}$ to get the
average tangle$\ A$.

One can see that $A=0$ for the GHZ state and $A=4/9$ for the W state. It is
known that the W state has the maximal average tangles $A=4/9$ \cite{Dur}.
Thus, $0\leq A\leq 4/9$.

\subsection{Definition for the average von Neumann entropy}

Let $m(\psi)$ be the average of the von Neumann entropy of all the reduced
density operators for the state $|\psi \rangle $. Then, 
\begin{equation}
m(\psi)=\frac{S(\rho _{A})+S(\rho _{B})+S(\rho _{C})}{3}.  \label{average-2}
\end{equation}

We explain how to calculate the average von Neumann entropy $m$. First, take
partial traces over qubits A and B (resp. A and C, and B and C) to get\ the
reduced density operators $\rho _{C}$ (resp. $\rho _{B}$, and $\rho _{A}$),
then calculate the von Neumann entropy $S(\rho _{A})$ (resp. $S(\rho _{B})$,
and $S(\rho _{C})$) of $\rho _{A}$(resp. $\rho _{B}$, and $\rho _{C}$) by
the definition in Eq. (\ref{von-1}). Finally, make the average of $S(\rho
_{A})$, $S(\rho _{B})$, and $S(\rho _{C}))$ to get the average von Neumann
entropy $m$.

%
%
%
%
%
%
%
%
%
%
%
%
%
%
%
%
%
%
%
%
It is easy to see that the GHZ state has the maximal average von Neumann
entropy of $\ln 2$, while the W state has the average von Neumann entropy of 
$\displaystyle\frac{3\ln 3-2\ln 2}{3}$.%
%
%
%
%
%
%
%
%
%
%
Thus, $0\leq m\leq \ln 2$.

\subsection{Relation among the average tangle, the average von Neumann
entropy, and 3-tangle}

So far, no one explains why the GHZ\ state has the maximal 3-tangle but
vanishing tangle and conversely, and why the W state has the maximal average
tangle but vanishing 3-tangle. We will answer why the states GHZ and W have
the opposite properties.

Using Eq. (\ref{Taylor-1}), the average von Neumann entropy and the average
tangle satisfy\ the following equation for any pure state of three qubits. 
\begin{equation}
m-A-\frac{\tau _{ABC}}{2}\approx \ln 2-\frac{1}{2},  \label{relat-0}
\end{equation}%
where $0\leq m\leq \ln 2$ and $0\leq A\leq 4/9$.

Eq. (\ref{relat-0}) reveals a relation among the average von Neumann entropy 
$m$, the average tangle $A$, and the 3-tangle $\tau _{ABC}$. Clearly, $m=\ln
2$, $A=4/9$, $\tau _{ABC}$ $=1$ don't satisfy Eq. (\ref{relat-0}). It means
that for any state, the average von Neumann entropy, the average tangle, and
3-tangle cannot reach the maximum simultaneously. Below, we will investigate
when $m$ (resp. $A$ and $\tau _{ABC}$) reaches the maximum, what happen to
other two measures.

Equation (\ref{relat-0}) implies that the value of $(m-A)$, i.e. the
difference between the average von Neumann entropy and the average tangle,
increases linearly with the 3-tangle $\tau _{ABC}$. For the GHZ\ SLOCC
class, $0<\tau _{ABC}\leq 1$ and almost $\ln 2-1/2<m-A\leq \ln 2$. While for
other SLOCC\ classes, $\tau _{ABC}=0$ and $(m-A)\approx \ln 2-1/2$. Thus, we
obtain almost $\ln 2-1/2\leq (m-A)\leq \ln 2$ for any state of three qubits.

\subsection{The relation can help find the states which satisfy different
requirements for the average von Neumann entropy and the average tangle and
3-tangle.}

Clearly, the requirements for the average tangle, the average von Neumann
entropy, and 3-tangle must satisfy Eq. (\ref{relat-0}). For example, $%
|G\rangle $ in Table III has $A=1/4$, $m=0.56$, and $\tau _{ABC}=1/4$.

It is known that the GHZ state has vanishing tangle. This means that for the
GHZ\ state, if one of three qubits is traced out, the corresponding reduced
density operator becomes separable. In other words, the entanglement
properties of the GHZ state are fragile under particle losses \cite{Dur}.
One can use Eq. (\ref{relat-0}) to find the states of which the average von
Neumann entropy, the average tangle, and 3-tangle are big enough. Clearly,
the states are genuine entangled ones even losing one particle.

\subsection{The relation helps understand properties of the average tangle,
3-tangle, and the average von Neumann entropy}

We next explore when a state has the maximal average von Neumann entropy,
then what Eq. (\ref{relat-0}) can tell us about the average tangle and the
3-tangle for this state.

Let $m=\ln 2$ (for example, for the GHZ state). Then, Eq. (\ref{relat-0})
becomes 
\begin{equation}
A+\frac{\tau _{ABC}}{2}\approx \frac{1}{2}.  \label{relat-1}
\end{equation}%
Clearly, if $A=0$ then $\tau _{ABC}\approx 1$, and vice versa. Eq. (\ref%
{relat-1}) tells us if a state has the maximal average von Neumann entropy,
then the average tangle and the 3-tangle must satisfy Eq. (\ref{relat-1}).

For example, Eq. (\ref{relat-1}) tells us there are the states with $m=\ln 2$%
, $\tau _{ABC}\approx 1/2$, and $A\approx 1/4$. One can see that these
states belong to GHZ SLOCC\ class, but different from the states GHZ state.
The three entanglement measures: tangle, 3-tangle, and von Neumann entropy
tell us that the states are genuine entangled state even though one of three
qubits is traced over and have the maximal average von Neumann entropy.

We next explore when a state has the maximal 3-tangle, then what Eq. (\ref%
{relat-0}) can tell us about the average von Neumann entropy and the average
tangle for this state.

Let $\tau _{ABC}=1$ (for example, for the GHZ state). For the case, Eq. (\ref%
{relat-0}) becomes 
\begin{equation}
m-\ln 2\approx A.  \label{average}
\end{equation}%
It is easy\ to see that $m=\ln 2$ and $A=0$ is the unique solution of Eq. (%
\ref{average}) because if $m<\ln 2$ then $A<0$.

This explains when a state has the maximal 3-tangle, then the state must
have the maximal average von Neumann entropy and vanishing the average
tangle, i.e. $\tau _{\mu \nu }=0$, where $\mu \nu =\{AB,AC,BC\}$. This is
why the GHZ\ state has the maximal 3-tangle and the maximal average von
Neumann entropy but vanishing tangle $\tau _{\mu \nu }$, where $\mu \nu
=\{AB,AC,BC\}$.

We next explore when a state has the maximal average tangles of $4/9$, then
what Eq. (\ref{relat-0}) can tell us about the average von Neumann entropy
and the 3-tangle for this state.

Let $A=4/9$ (for example, for the W state). Via Eq. (\ref{relat-0}), one can
know that $\max \tau _{ABC}\approx 1/9$ when $m=\ln 2$ and $\min m\approx
\ln 2-1/18=\allowbreak 0.637\,59$ when $\tau _{ABC}=0$. It means the average
von Neumann entropy almost is maximal.

This explains when a state has the maximal average tangle of $4/9$, then the
state has almost vanishing 3-tangle and the almost maximal average von
Neumann entropy. This is why the W state has the maximal average tangles but
vanishing 3-tangle.

\vspace{10pt}

\section{Tangles and von Neumann entropy for GHZ SLOCC\ class}

It is well known that an ASD state belongs to the GHZ$\ $SLOCC\ class 
\textit{iff} $\lambda _{0}\lambda _{4}\neq 0$ \cite{Dli-qip-18, Dli-jpa-20}.
So, in this section we assume $\lambda _{0}\lambda _{4}\neq 0$. We first
discuss the properties of tangle. 
It is known that if $\tau _{\mu \nu }$, vanishes, then $\rho _{\mu \nu }$ is
separable, where $\mu \nu \in \{AB,AC,BC\}$. From Table \ref{T1}, one can
see that the tangles for the GHZ SLOCC\ class have the following properties.

\vspace{10pt} \textit{Property (1).} $0\leq \tau _{AB}$, $\tau _{AC}$, $\tau
_{BC}<1$.

\vspace{10pt}

\textit{Property (2).}

(2.1). $\tau _{AB}=0$ \textit{iff} $\lambda _{3}=0$.

(2.2). $\tau _{AC}=0$ \textit{iff} $\lambda _{2}=0$.

(2.3). $\tau _{BC}=0$ \textit{iff} $\lambda _{1}\lambda _{4}=\lambda
_{2}\lambda _{3}\neq 0$ and $\phi =0$ or $\lambda _{1}=0$ and $\lambda
_{2}\lambda _{3}=0 $.

\vspace{10pt}

\textit{Property (3).} When only one of $\tau _{AB}$, $\tau _{AC}$, and $%
\tau _{BC}$ vanishes,

(3.1). $\tau _{AB}=0$ and $\tau _{AC}\tau _{BC}\neq 0$ \textit{iff} $\lambda
_{3}=0$ and $\lambda _{1}\lambda _{2}\neq 0$.

(3.2). $\tau _{AC}=0$ and $\tau _{AB}\tau _{BC}\neq 0$ \textit{iff} $\lambda
_{2}=0$ and $\lambda _{1}\lambda _{3}\neq 0$.

(3.3). $\tau _{BC}=0$ and $\tau _{AB}\tau _{AC}\neq 0$ \textit{iff} $\lambda
_{1}\lambda _{4}=\lambda _{2}\lambda _{3}\neq 0$ and $\phi =0$.

\vspace{10pt}

{\underline {Proof of (3.3)}}: Clearly, $\tau _{AB}\tau _{AC}\neq 0$ \textit{%
iff} $\lambda _{2}\lambda _{3}\neq 0$. If $\tau _{BC}=0$, then from Table %
\ref{T1},$\ \lambda _{1}\lambda _{4}e^{i\phi }-\lambda _{2}\lambda _{3}=0$,
i.e. $\lambda _{1}\lambda _{4}=\lambda _{2}\lambda _{3}\neq 0$ and $\phi =0$%
. Conversely, if $\lambda _{1}\lambda _{4}=\lambda _{2}\lambda _{3}\neq 0$
and $\phi =0$, then $\tau _{AB}\tau _{AC}\neq 0$ and $\tau _{BC}=0$.

\vspace{10pt}

\textit{Property (4).} When only two of $\tau _{AB}$, $\tau _{AC}$, and $%
\tau _{BC}$ vanish,

(4.1). $\tau _{AB}=\tau _{AC}=0$ and $\tau _{BC}\neq 0$ iff $\lambda
_{2}=\lambda _{3}=0$ and $\lambda _{1}\neq 0$.

(4.2). $\tau _{AB}=\tau _{BC}=0$ and $\tau _{AC}\neq 0$ iff \ $\lambda
_{1}=\lambda _{3}=0$\ \ and $\lambda _{2}\neq 0$.

(4.3). $\tau _{AC}=\tau _{BC}=0$ and $\tau _{AB}\neq 0$ iff $\lambda
_{1}=\lambda _{2}=0$\ \ \ and $\lambda _{3}\neq 0$.

\vspace{10pt}

\textit{Property (5).} When all $\tau _{AB}$, $\tau _{AC}$, and $\tau _{BC}$
vanish,

$\tau _{AB}=\tau _{AC}=\tau _{BC}=0$ iff \ $\lambda _{1}=\lambda
_{2}=\lambda _{3}=0$, i.e. the state is of the form $\lambda _{0}|000\rangle
+\lambda _{4}|111\rangle $. For instance, the GHZ state.

\vspace{10pt}

\textit{Property (6).} When none of the tangles vanishes, $\tau _{AB}\tau
_{AC}\tau _{BC}>0$ iff $\ $(i). $\lambda _{2}\lambda _{3}\neq 0$ and $%
\lambda _{1}=0$, or (ii). $\lambda _{1}\lambda _{2}\lambda _{3}(\lambda
_{1}\lambda _{4}-\lambda _{2}\lambda _{3})\neq 0$ and $\phi =0$, or (iii). $%
\lambda _{1}\lambda _{2}\lambda _{3}\neq 0$ and $\phi \neq 0$.

\subsection{The GHZ state is the unique one under ASD which has the maximal
von Neumann entropy .}

In Appendix B, we show that for any ASD state of three qubits, if $S(\rho
_{\mu })=\ln 2$, then the state must be GHZ. It means that the GHZ\ state is
the unique state of three qubits under ASD such that $S(\rho _{\mu })$
achieves the maximal value. Thus, for any state of GHZ LU class, $S(\rho
_{\mu })=\ln 2$, where $\mu \in \{A,B,C\}$.

\section{Tangle and von Neumann entropy for W SLOCC\ class}

Each state of the W SLOCC\ class is of the following form \cite{Dli-qip-18,
Dli-jpa-20}, 
\begin{equation}
|\psi \rangle =\lambda _{0}|000\rangle +\lambda _{1}e^{i\phi }|100\rangle
+\lambda _{2}|101\rangle +\lambda _{3}|110\rangle ,  \label{w-0}
\end{equation}%
where $\lambda _{0}\lambda _{2}\lambda _{3}\neq 0$ and $\lambda _{4}=0$. \ 

\subsection{Tangle for the W SLOCC\ class}

From Table \ref{T1}, we obtain tangles for the W SLOCC\ class, 
\begin{eqnarray*}
\tau _{AB} &=&4\lambda _{0}^{2}\lambda _{3}^{2}, \\
\tau _{AC} &=&4\lambda_{0}^{2}\lambda _{2}^{2}, \\
\tau _{BC} &=&4\lambda _{2}^{2}\lambda _{3}^{2}, \\
\tau _{ABC} &=&0.
\end{eqnarray*}
Clearly, $0<\tau _{AB},\tau _{AC},\tau _{BC}<1$.

\subsection{The W state is the unique state of the W SLOCC\ class under ASD
whose average von Neumann entropy is the maximal within the W SLOCC\ class.}

\vspace{10pt}

Next, we show that the average von Neumann entropy of the W state is maximal
only within the W SLOCC\ class. From Table \ref{T2}, for W SLOCC class, one
can see that 
\begin{eqnarray}
\alpha _{A} &=&\lambda _{0}^{2}(\lambda _{2}^{2}+\lambda _{3}^{2}),
\label{wsc-1} \\
\alpha _{B} &=&\lambda _{3}^{2}(\lambda _{0}^{2}+\lambda _{2}^{2}),
\label{wsc-2} \\
\alpha _{C} &=&\lambda _{2}^{2}(\lambda _{0}^{2}+\lambda _{3}^{2}).
\label{wsc-3}
\end{eqnarray}

Let $\mu \nu \upsilon $ be a string from the set $\{ABC,BAC,CAB\}$. When $%
\alpha _{\mu }=1/4$, $S(\rho _{\mu })=\ln 2$, $S(\rho _{\nu })<\ln 2$ and $%
S(\rho _{\upsilon })<\ln 2$. It means that the von Neumann entropy of the W
state is not maximal even within the W SLOCC\ class. We next show that the
average von Neumann entropy of the W state is maximal within the W SLOCC\
class.

\vspace{10pt} In Appendix C, we derive the extrema of $m$ with the
constraint $\sum_{i=0}^{3}\lambda _{i}^{2}=1$. A straightforward and tedious
calculation yields that $m$ has the extrema$\frac{3\ln 3-2\ln 2}{3}$ at $%
\lambda _{1}=0$ and $\lambda _{0}=\lambda _{2}=\lambda _{3}=1/\sqrt{3}$,
which is just the ASD\ of the W state, and the extrema is the maximum.
Therefore, the W state is the unique state under ASD\ of which the average
von Neumann entropy of all kinds of the reduced density operators is maximal
within the W SLOCC class, although the W state does not have the maximal
average of the von Neumann entropy for all states of three qubits.

\section{The states for which the tangle, concurrence, 3-tangle and von
Neumann entropy don't vanish}

It is known that $\tau _{ABC}$ vanishes for the SLOCC\ classes W, A-BC,
B-AC, C-AB, and A-B-C and the tangles $\tau _{AB}\geq 0,\tau _{AC}\geq 0,$
and $\tau _{BC}\geq 0$ for the GHZ\ SLOCC\ class. Next, we present all the
states for which the tangles, concurrence, 3-tangle and von Neumann entropy
do not vanish. We know these states can only come from the GHZ SLOCC\ class.
For these states, when one of three qubits is traced out, their reduced
density operators are entangled while the reduced density operator of the
GHZ state is separable. It seems that the states are more entangled than the
GHZ state under tangle and the W state under 3-tangle, respectively. Note
that the\ tangle is the square of the concurrence. Therefore, if a tangle
does not vanish, the concurrence does not either, and hence, we do not need
to discuss the concurrence separately.

\vspace{10pt}

Therefore, if a state satisfies $\tau _{AB}\tau _{AC}\tau _{BC}\neq 0$, $%
S(\rho _{A})S(\rho _{B})S(\rho _{C})\neq 0$, and $\tau _{ABC}\neq 0$, then
the state belongs to the GHZ SLOCC\ class and it falls in one of the three
cases: $\ $(i). $\lambda _{2}\lambda _{3}\neq 0$ and $\lambda _{1}=0$, i.e.,
the state is 
\begin{equation}
|\varpi _{1}\rangle =\lambda _{0}|000\rangle +\lambda _{2}|101\rangle
+\lambda _{3}|110\rangle +\lambda _{4}|111\rangle ,  \label{full-e-1}
\end{equation}%
or (ii). $\lambda _{1}\lambda _{2}\lambda _{3}(\lambda _{1}\lambda
_{4}-\lambda _{2}\lambda _{3})\neq 0$ and $\phi =0$, i.e., the state is 
\begin{equation}
|\varpi _{2}\rangle =\lambda _{0}|000\rangle +\lambda _{1}|100\rangle
+\lambda _{2}|101\rangle +\lambda _{3}|110\rangle +\lambda _{4}|111\rangle ,
\label{full-e-2}
\end{equation}%
where $\lambda _{1}\lambda _{4}\neq \lambda _{2}\lambda _{3}$, \ or (iii). $%
\lambda _{1}\lambda _{2}\lambda _{3}\neq 0$ and $\phi \neq 0$, i.e., the
state is 
\begin{eqnarray}
|\varpi _{3}\rangle &=&\lambda _{0}|000\rangle +\lambda _{1}e^{i\phi
}|100\rangle +\lambda _{2}|101\rangle  \notag \\
&&+\lambda _{3}|110\rangle +\lambda _{4}|111\rangle.  \label{full-e-3}
\end{eqnarray}

Thus, all states with non-vanishing tangles, 3-tangle, and von Neumann
entropy can be written in the form of $|\varpi _{1}\rangle $, $|\varpi
_{2}\rangle $ or $|\varpi _{3}\rangle $ with suitable $\lambda $'s. For
example, the following states in the form of $|\varpi _{1}\rangle $. 
\begin{eqnarray}
|\kappa \rangle &=&\frac{2}{3}|000\rangle +\frac{1}{2}|101\rangle +\frac{1}{2%
}|110\rangle +\frac{\sqrt{2}}{6}|111\rangle , \\
|G\rangle &=&\frac{1}{2}(|000\rangle +|101\rangle +|110\rangle +|111\rangle
), \\
|\vartheta \rangle &=&\frac{\sqrt{5}}{10}(3|000\rangle +|101\rangle
+|110\rangle +3|111\rangle ).
\end{eqnarray}%
See Table \ref{T5} for details. Furthermore, there exist other interesting
states approximating the tangle and 3-tangle of the W state and the GHZ
state.

\vspace{10pt} (a). Let $\lambda_2=\lambda_3=\lambda_0$ in $|\varpi
_{1}\rangle $, we obtain a new state 
\begin{equation*}
|\omega \rangle =\lambda _{0}|000\rangle +\lambda _{0}|101\rangle +\lambda
_{0}|110\rangle +\lambda _{4}|111\rangle .
\end{equation*}

The tangles are 
\begin{equation}
\tau _{AB}=\tau _{AC}=\tau _{BC}=A=4\lambda _{0}^{4}=\frac{4}{9}(1-\lambda
_{4}^{2})^{2}.  \label{omiga-1}
\end{equation}%
From Eq. (\ref{omiga-1}), we know that when $\lambda _{4}$ is small enough,
then the tangles, the average tangle, 3-tangle, and von Neumann entropy of $%
|\omega \rangle $ are almost equal to those of the W state.

\vspace{10pt}

(b). Let $\lambda_4=\lambda_0$, and let $\lambda_3=\lambda_2$, we obtain a
new state 
\begin{equation*}
|\varkappa \rangle =\lambda _{0}|000\rangle +\lambda _{2}|101\rangle
+\lambda _{2}|110\rangle +\lambda _{0}|111\rangle .
\end{equation*}%
The tangles and 3-tangle are given by 
\begin{equation}
\tau _{AB}=\tau _{AC}=4\lambda _{0}^{2}\lambda _{2}^{2},\tau _{BC}=4\lambda
_{2}^{4},\tau _{ABC}=4\lambda _{0}^{4}.
\end{equation}

For $|\varkappa \rangle $, we have $\lim_{\lambda _{2}\rightarrow 0}S(\rho
_{\mu })=\ln 2$, where $\mu \in \{A,B,C\}$, $\lim_{\lambda _{2}\rightarrow
0}\tau _{ABC}=1$, and $\lim_{\lambda _{2}\rightarrow 0}\tau _{\mu \upsilon
}=0$, where $\mu \upsilon \in \{AB,AC,BC\}$. Thus, when$\ \lambda _{2}$ is
small enough, then the tangles, the average tangle, 3-tangle, and von
Neumann entropy of $|\varkappa \rangle $ are almost equal to those of the
GHZ state. See $|\vartheta \rangle $ in Table \ref{T5}.

\vspace{10pt}

Next, we demonstrate that tangles $\tau _{AB}$, $\tau _{AC}$, and $\tau
_{BC} $ with $\tau _{AB}\tau _{AC}\tau _{BC}\neq 0$ determine a unique state
of the form of $|\varpi _{1}\rangle $. In other words, the LU invariants $%
\{J_i | i=1,2,3\}$ with $J_{1}J_{2}J_{3}\neq 0$ can determine a unique state
of the form of $|\varpi _{1}\rangle $.

\vspace{10pt}

Suppose the tangles $\tau _{AB}=4\lambda _{0}^{2}\lambda _{3}^{2}=p^{4}$, $%
\tau _{AC}=4\lambda _{0}^{2}\lambda _{2}^{2}=q^{4}$, and $\tau
_{BC}=4\lambda _{2}^{2}\lambda _{3}^{2}=r^{4}$, where $pqr\neq 0$. It is
known that $p^{4}+q^{4}+r^{4}<4/3$ from the CKW\ inequality $\tau _{AB}+\tau
_{AC}+\tau _{BC}<4/3$ (see Appendix B and also \cite{Dur}). The unique state
corresponding to the tangles is 
\begin{equation*}
\frac{1}{\sqrt{2}}\frac{pq}{r}|000\rangle +\frac{1}{\sqrt{2}}\frac{qr}{p}%
|101\rangle +\frac{1}{\sqrt{2}}\frac{pr}{q}|110\rangle +\lambda
_{4}|111\rangle,
\end{equation*}
where $\lambda _{4}^{2}=1-\frac{1}{2}((\frac{pq}{r})^{2}+(\frac{qr}{p})^{2}+(%
\frac{pr}{q})^{2})$.

\vspace{10pt} On the other hand, the five LU invariants $\{J_i | i=1, \ldots
5\}$ cannot uniquely determine a state of three qubits. For instance, for $%
|\varpi _{1}\rangle $, from that $\lambda _{0}^{2}+\sum_{i=2}^{4}\lambda
_{i}^{2}=1$ we can derive $J_{4}=\sqrt{J_{2}J_{3}/J_{1}}%
-(J_{2}J_{3}/J_{1}+J_{2}+J_{3})$. That is, $J_{4}$ is a function of $J_{1}$, 
$J_{2}$, and $J_{3}$, which explains why $\{J_{i}| i=1,\cdots ,5\}$ cannot
determine a unique state of three qubits \cite{Acin01}.

\begin{table}[ht]
\caption{Tangle, 3-tangle and von Neumann entropy for some states ($\protect%
\delta =\frac{3\ln 3-2\ln 2}{3}=\allowbreak 0.63651$)}
\label{T5}\centering
\begin{tabular}{|l|l|l|l|l|l|l|l|}
\toprule &  &  &  &  &  &  &  \\ 
State & $\tau _{AB}$ & $\tau _{AC}$ & $\tau _{BC}$ & $S(\rho _{A})$ & $%
S(\rho _{B})$ & $S(\rho _{C})$ & $\tau _{ABC}$ \\ 
&  &  &  &  &  &  &  \\ 
\colrule &  &  &  &  &  &  &  \\ 
GHZ & $0$ & $0$ & $0$ & $\ln 2$ & $\ln 2$ & $\ln 2$ & $1$ \\ 
&  &  &  &  &  &  &  \\ 
\colrule &  &  &  &  &  &  &  \\ 
W & $\frac{4}{9}$ & $\frac{4}{9}$ & $\frac{4}{9}$ & $\delta $ & $\delta $ & $%
\delta $ & $0$ \\ 
&  &  &  &  &  &  &  \\ 
\colrule &  &  &  &  &  &  &  \\ 
$|G\rangle $ & $\frac{1}{4}$ & $\frac{1}{4}$ & $\frac{1}{4}$ & $0.56$ & $%
0.56 $ & $0.56$ & $\frac{1}{4}$ \\ 
&  &  &  &  &  &  &  \\ 
\colrule &  &  &  &  &  &  &  \\ 
$|\kappa \rangle $ & $\frac{4}{9}$ & $\frac{4}{9}$ & $\frac{1}{4}$ & $0.687$
& $0.587$ & $0.587$ & $\frac{8}{81}$ \\ 
&  &  &  &  &  &  &  \\ 
\colrule &  &  &  &  &  &  &  \\ 
$|\vartheta \rangle $ & $\frac{9}{100}$ & $\frac{9}{100}$ & $\frac{1}{100}$
& $0.688$ & $\allowbreak 0.647$ & $0.647$ & $\frac{81}{100}$ \\ 
&  &  &  &  &  &  &  \\ 
\botrule &  &  &  &  &  &  & 
\end{tabular}%
\end{table}

\section{CKW inequalities for GHZ SLOCC\ class}

For GHZ SLOCC\ class, $\lambda _{0}\lambda _{4}\neq 0$, we have the
following CKW inequalities. The arguments for CKW inequalities are
summarized in Appendix D.

\begin{enumerate}
\item If $\lambda _{1}=\lambda _{2}=\lambda _{3}=0$, then $\tau _{AB}+\tau
_{AC}+\tau _{BC}=0$. This is the generated GHZ\ state $p|000\rangle
+q|111\rangle $.

\item If $\lambda _{1}=\lambda _{2}=0$, and $\lambda _{3}\neq 0$, then $\tau
_{AB}+\tau _{AC}+\tau _{BC}<1$.

\item If $\ \lambda _{1}=\lambda _{3}=0$, and $\lambda _{2}\neq 0$,$\ $ then 
$\tau _{AB}+\tau _{AC}+\tau _{BC}<1$.

\item If $\lambda _{2}=\lambda _{3}=0$, and $\lambda _{1}\neq 0$, then $\tau
_{AB}+\tau _{AC}+\tau _{BC}<1$.

\item If $\lambda _{1}=0$, and $\lambda _{2}\lambda _{3}\neq 0$, then $\tau
_{AB}+\tau _{AC}+\tau _{BC}<\frac{4}{3}$.

\item If $\lambda _{2}=0$, and $\lambda _{1}\lambda _{3}\neq 0$, then $\tau
_{AB}+\tau _{AC}+\tau _{BC}\leq \frac{1}{2}$.

\item If $\lambda _{3}=0$, and $\lambda _{1}\lambda _{2}\neq 0$, then $\tau
_{AB}+\tau _{AC}+\tau _{BC}\leq \frac{1}{2}$.

\item If $\lambda _{1}\lambda _{2}\lambda _{3}\neq 0$, then $\tau _{AB}+\tau
_{AC}+\tau _{BC}\leq 1$.
\end{enumerate}

\section{Summary}

In this paper, we have given a general formula of the tangles for pure
states of three qubits, which are LU polynomial invariants of degree 4. We
derived tangles and von Neumann entropy for ASD, and presented a relation
among tangles, 3-tangle, and von Neumann entropy , as well as a relation
among the average tangle, the average von Neumann entropy , and 3-tangle.

\vspace{10pt}

Via the von Neumann entropy for ASD, we indicated that the GHZ state is the
unique state of three qubits under ASD that has the maximal von Neumann
entropy for all kinds of the reduced density operators, while the average
von Neumann entropy of the W state is maximal only within the W SLOCC\ class.

\vspace{10pt} It is known that the tangles of the GHZ state and the 3-tangle
of the W state would vanish. We obtained all states of three qubits with
non-vanishing tangle, concurrence, 3-tangle and von Neumann entropy . For
example, $|\vartheta \rangle $ is such a state. It means that when one of
the three qubits is traced out, the remaining state from $|\vartheta \rangle 
$ is still entangled, while the remaining state of the GHZ state becomes
separable. From Table \ref{T5}, it is shown that $S(\rho _{\mu })$ of $%
|\vartheta \rangle $ is bigger than that of the W state, where $\mu \in
\{A,B,C\}$. Therefore, state $|\vartheta \rangle $ seems to be more
entangled than the GHZ\ state under tangle and more tangled than the W state
under von Neumann entropy and 3-tangle.

\section{Appendix A. Calculation of tangles.}

\setcounter{equation}{0} \renewcommand{\theequation}{A\arabic{equation}} (A)
Calculating $\tau _{AB}$

\vspace{10pt} From the definition of $\tau _{AB}$ in Eq. (\ref{tangle-1}),
we obtain the characteristic polynomial (CP) of $\rho _{AB}\overline{\rho
_{AB}}$ as follows, 
\begin{equation}
(X^{2}-\Delta X+\Theta ^{2})X^{2},
\end{equation}%
where 
\begin{eqnarray}
\Theta&=&|(c_{0}c_{7}-c_{2}c_{5})^{2}+(c_{1}c_{6}-c_{3}c_{4})^{2}  \notag \\
&&-2(c_{0}c_{7}+c_{2}c_{5})(c_{1}c_{6}+c_{3}c_{4})  \notag \\
&&+4c_{0}c_{3}c_{5}c_{6}+4c_{1}c_{2}c_{4}c_{7}|,
\end{eqnarray}

\begin{eqnarray}
\Delta &=&2\left( \left\vert c_{0}\right\vert ^{2}+\left\vert
c_{1}\right\vert ^{2}\right) \left( \left\vert c_{6}\right\vert
^{2}+\left\vert c_{7}\right\vert ^{2}\right)  \notag \\
&&+2\left( \left\vert c_{2}\right\vert ^{2}+\left\vert c_{3}\right\vert
^{2}\right) \left( \left\vert c_{4}\right\vert ^{2}+\left\vert
c_{5}\right\vert ^{2}\right)  \notag \\
&&+2\left\vert c_{0}c_{6}^{\ast }+c_{1}c_{7}^{\ast }\right\vert
^{2}+2\left\vert c_{2}c_{4}^{\ast }+c_{3}c_{5}^{\ast }\right\vert ^{2} 
\notag \\
&&-4\ast re(\left( c_{0}c_{2}^{\ast }+c_{1}c_{3}^{\ast }\right) \left(
c_{6}c_{4}^{\ast }+c_{7}c_{5}^{\ast }\right) )  \notag \\
&&-4\ast re(\left( c_{0}c_{4}^{\ast }+c_{1}c_{5}^{\ast }\right) \left(
c_{6}c_{2}^{\ast }+c_{7}c_{3}^{\ast }\right) ),  \label{tang-ab}
\end{eqnarray}%
and $re(c)$ indicates the real part of a complex number $c$. Moreover, one
can find 
\begin{equation}
4\Theta =\tau _{ABC}\text{.}
\end{equation}%
Hence, the eigenvalues of $\rho _{AB}\overline{\rho _{AB}}$ are $0,0,$ and $%
\frac{\Delta \pm \sqrt{\Delta ^{2}-4\Theta ^{2}}}{2}$. It is known that $%
\rho _{AB}\overline{\rho _{AB}}$ has only real and non-negative eigenvalues 
\cite{Coffman}. Then, by the definition of $\tau _{AB}$ in Eq. (\ref%
{tangle-1}), we obtain 
\begin{eqnarray}
\tau _{AB} &=&\left( \sqrt{\frac{\Delta +\sqrt{\Delta ^{2}-4\Theta ^{2}}}{2}}%
-\sqrt{\frac{\Delta -\sqrt{\Delta ^{2}-4\Theta ^{2}}}{2}}\right) ^{2}  \notag
\\
&=&\Delta -2\Theta  \notag \\
&=&\Delta -\frac{\tau _{ABC}}{2}.  \label{poly-1}
\end{eqnarray}

(B) Calculating $\tau _{AC}$

\vspace{10pt}

Similarly, we can obtain the CP of $\rho _{AC}\overline{\rho _{AC}}$ as
follows, 
\begin{equation}
X^{2}(X^{2}-\Phi X+\Upsilon ^{2}),
\end{equation}%
where

\begin{equation}
\Upsilon =\Theta ,
\end{equation}

\begin{eqnarray}
\Phi &=&2\left( \left\vert c_{0}\right\vert ^{2}+\left\vert c_{2}\right\vert
^{2}\right) \left( \left\vert c_{5}\right\vert ^{2}+\left\vert
c_{7}\right\vert ^{2}\right)  \notag \\
&&+2\left( \left\vert c_{1}\right\vert ^{2}+\left\vert c_{3}\right\vert
^{2}\right) \left( \left\vert c_{4}\right\vert ^{2}+\left\vert
c_{6}\right\vert ^{2}\right)  \notag \\
&&+2\left\vert c_{0}c_{5}^{\ast }+c_{2}c_{7}^{\ast }\right\vert
^{2}+2\left\vert c_{1}c_{4}^{\ast }+c_{3}c_{6}^{\ast }\right\vert ^{2} 
\notag \\
&&-4\ast re(\left( c_{0}c_{1}^{\ast }+c_{2}c_{3}^{\ast }\right) \left(
c_{5}c_{4}^{\ast }+c_{7}c_{6}^{\ast }\right) )  \notag \\
&&-4\ast re(\left( c_{0}c_{4}^{\ast }+c_{2}c_{6}^{\ast }\right) \left(
c_{5}c_{1}^{\ast }+c_{7}c_{3}^{\ast }\right) ).  \notag \\
&&  \label{tang-ac}
\end{eqnarray}%
Then, we obtain 
\begin{equation}
\tau _{AC}=\Phi -2\Upsilon =\Phi -\frac{\tau _{ABC}}{2}.  \label{poly-2}
\end{equation}

(C) Calculating $\tau _{BC}$

\vspace{10pt}

Similarly, the characteristic polynomial of $\rho _{BC}\overline{\rho _{BC}}$
is given by 
\begin{equation}
X^{2}(X^{2}-\Psi X+\digamma ^{2}),
\end{equation}%
where \ 
\begin{equation}
\digamma =\Theta ,
\end{equation}%
\ 
\begin{eqnarray}
\Psi &=&2\left( \left\vert c_{0}\right\vert ^{2}+\left\vert c_{4}\right\vert
^{2}\right) \left( \left\vert c_{3}\right\vert ^{2}+\left\vert
c_{7}\right\vert ^{2}\right)  \notag \\
&&+2\left( \left\vert c_{1}\right\vert ^{2}+\left\vert c_{5}\right\vert
^{2}\right) \left( \left\vert c_{2}\right\vert ^{2}+\left\vert
c_{6}\right\vert ^{2}\right)  \notag \\
&&+2\left\vert c_{0}c_{3}^{\ast }+c_{4}c_{7}^{\ast }\right\vert
^{2}+2\left\vert c_{1}c_{2}^{\ast }+c_{5}c_{6}^{\ast }\right\vert ^{2} 
\notag \\
&&-4\ast re(\left( c_{0}c_{1}^{\ast }+c_{4}c_{5}^{\ast }\right) \left(
c_{3}c_{2}^{\ast }+c_{7}c_{6}^{\ast }\right) )  \notag \\
&&-4\ast re(\left( c_{0}c_{2}^{\ast }+c_{4}c_{6}^{\ast }\right) \left(
c_{3}c_{1}^{\ast }+c_{7}c_{5}^{\ast }\right) ).  \label{tang-bc}
\end{eqnarray}%
Then, we obtain 
\begin{equation}
\tau _{BC}=\Psi -2\digamma =\Psi -\frac{\tau _{ABC}}{2}.  \label{poly-3}
\end{equation}

(D) By solving CKW equations \cite{Coffman} 
\begin{eqnarray}
\tau _{AB}+\tau _{AC}+\tau _{ABC} &=&\tau _{A(BC)}, \\
\tau _{AB}+\tau _{BC}+\tau _{ABC} &=&\tau _{B(AC)}, \\
\tau _{AC}+\tau _{BC}+\tau _{ABC} &=&\tau _{C(AB)},
\end{eqnarray}%
where $\tau _{A(BC)}=4\det \rho _{A}$, $\tau _{B(AC)}=4\det \rho _{B}$, and $%
\tau _{C(AB)}=4\det \rho _{C}$, we obtain 
\begin{eqnarray}
\tau _{AB} =&\frac{\tau _{A(BC)}+\tau _{B(AC)}-\tau _{C(AB)}-\tau _{ABC}}{2},
\\
\tau _{AC} =&\frac{\tau _{A(BC)}-\tau _{B(AC)}+\tau _{C(AB)}-\tau _{ABC}}{2},
\\
\tau _{BC} =&\frac{-\tau _{A(BC)}+\tau _{B(AC)}+\tau _{C(AB)}-\tau _{ABC}}{2}%
.
\end{eqnarray}
When the 3-tangle $\tau_{ABC}$ is zero, we obtain the following, 
\begin{eqnarray}
\Delta &=&\frac{\tau _{A(BC)}+\tau _{B(AC)}-\tau _{C(AB)}}{2}, \\
\Phi &=&\frac{\tau _{A(BC)}-\tau _{B(AC)}+\tau _{C(AB)}}{2}, \\
\Psi &=&\frac{-\tau _{A(BC)}+\tau _{B(AC)}+\tau _{C(AB)}}{2}.
\end{eqnarray}

Obviously, $\Delta $, $\Phi $, $\Psi $ are simple polynomial of degree 4
although it is hard to compute $\det \rho _{A}$, $\det \rho _{B}$, and $\det
\rho _{C}$.

\section*{Appendix B The GHZ\ state is unique state of three qubits which
has maximally von Neumann entropy}

\setcounter{equation}{0} \renewcommand{\theequation}{B\arabic{equation}}

{\underline{Claim:}} If a state of three qubits possesses the maximal von
Neumann entropy , $S(\rho _{\mu })=\ln 2$, where $\mu \in \{A,B,C\}$, then
the state must be GHZ.

\vspace{10pt}

{\underline {Proof:}} Clearly, $S(\rho _{\mu })$ increases strictly
monotonically as $\alpha _{\mu }$ increases. Therefore, $S(\rho _{\mu })=\ln
2$ \textit{iff} $\alpha _{\mu }=1/4$. Thus, we have the following equations 
\begin{eqnarray}
\alpha _{A} &=&J_{2}+J_{3}+J_{4}=1/4,  \label{eq-1} \\
\alpha _{B} &=&J_{1}+J_{3}+J_{4}=1/4, \\
\alpha _{C} &=&J_{1}+J_{2}+J_{4}=1/4,
\end{eqnarray}
and we obtain 
\begin{equation}
J_{1}=J_{2}=J_{3}.  \label{eq-4}
\end{equation}%
Using Tables \ref{T1} and \ref{T2}, equation (\ref{eq-1}) leads to 
\begin{equation}
\lambda _{0}^{4}-\lambda _{0}^{2}(1-\lambda _{1}^{2})+1/4=0.  \label{ghz-2}
\end{equation}
From (\ref{ghz-2}), we have a solution 
\begin{eqnarray}
\lambda _{1} &=&0,  \label{cod-1} \\
\lambda _{0} &=&1/\sqrt{2},  \label{cod-2} \\
\lambda _{2}^{2}+\lambda _{3}^{2}+\lambda _{4}^{2} &=&1/2.  \label{cod-3}
\end{eqnarray}

Using $J_{2}=J_{3}$\ in Eq. (\ref{eq-4}), we have 
\begin{equation}
\lambda _{2}=\lambda _{3}.  \label{eq-5}
\end{equation}
From that $J_{1}=J_{2}$ in Eq. (\ref{eq-4}), we obtain 
\begin{equation}
\lambda _{2}\lambda _{3}=\lambda _{0}\lambda _{2}.  \label{eq-6}
\end{equation}

There are two scenarios for $\lambda _{2}$: $\lambda _{2}\neq 0$ and $%
\lambda _{2}=0$. The first scenario is impossible since if $\lambda _{2}\neq
0$, then from Eqs. (\ref{cod-2}, \ref{eq-5}, \ref{eq-6}), we obtain 
\begin{equation}
\lambda _{0}=\lambda _{2}=\lambda _{3}=1/\sqrt{2}.  \label{ghz-3}
\end{equation}
From Eq. (\ref{cod-3}), we know clearly that Eq. (\ref{ghz-3}) cannot hold.
Therefore, $\lambda _{2}$ must be zero.

With $\lambda_2=0$, from Eqs. (\ref{cod-3}, \ref{eq-5}) we obtain 
\begin{eqnarray}
\lambda _{2} &=&\lambda _{3}=0,  \label{ghz-4} \\
\lambda _{4} &=&1/\sqrt{2}.  \label{ghz-5}
\end{eqnarray}
The state satisfying Eqs. (\ref{cod-1}, \ref{cod-2}, \ref{ghz-4}, \ref{ghz-5}%
) is just GHZ.

\section*{Appendix C The extrema for W SLOCC\ class}

\setcounter{equation}{0} \renewcommand{\theequation}{C\arabic{equation}}

We next find an extrema of $m$ with the constraint of $\sum_{i=0}^{3}\lambda
_{i}^{2}=1$ for the W SLOCC\ class. For states of W SLOCC\ class, $\alpha
_{A}=\lambda _{0}^{2}(\lambda _{2}^{2}+\lambda _{3}^{2})$, $\alpha
_{B}=\lambda _{3}^{2}(\lambda _{0}^{2}+\lambda _{2}^{2})$, and $\alpha
_{C}=\lambda _{2}^{2}(\lambda _{0}^{2}+\lambda _{3}^{2})$

\vspace{10pt} We define 
\begin{equation}
F=\frac{1}{3}(S(\rho _{A})+S(\rho _{B})+S(\rho _{C}))+\ell
(\sum_{i=0}^{3}\lambda _{i}^{2}-1),
\end{equation}%
where $\ell$ is the Lagrange multiplier. In light of the constrained extreme
theorem, we need to solve the equations $\frac{\partial F}{\partial \lambda
_{i}}=0$, for $i=0,1,2,3$, and $\frac{\partial F}{\partial \ell }=0$ to find
the extrema. From $\frac{\partial F}{\partial \lambda _{1}}=2\ell \lambda
_{1}=0$, we obtain $\lambda _{1}=0$. Then, $F$ is reduced to 
\begin{equation}
F=\frac{1}{3}(S(\rho _{A})+S(\rho _{B})+S(\rho _{C}))+\ell (\lambda
_{0}^{2}+\lambda _{2}^{2}+\lambda _{3}^{2}-1).  \label{ex-1}
\end{equation}%
From $\frac{\partial F}{\partial \ell }=0$ we obtain 
\begin{equation}
\lambda _{0}^{2}+\lambda _{2}^{2}+\lambda _{3}^{2}=1.  \label{const-1}
\end{equation}

From Eq. (\ref{ex-1}), \ we obtain 
\begin{equation}
\frac{\partial F}{\partial \lambda _{0}}=\frac{1}{3}\left[\frac{\partial
S(\rho _{A})}{\partial \lambda _{0}}+\frac{\partial S(\rho _{B})}{\partial
\lambda _{0}}+\frac{\partial S(\rho _{C})}{\partial \lambda _{0}} \right]%
+2\ell \lambda _{0},  \label{ex-1-2}
\end{equation}%
where 
\begin{equation}
\frac{\partial S(\rho _{\mu })}{\partial \lambda _{0}}=\frac{dS(\rho _{\mu })%
}{d\alpha _{\mu }}\frac{\partial \alpha _{\mu }}{\partial \lambda _{0}}, \mu
\in \{A,B,C\}.  \label{der-1}
\end{equation}

The derivative of $S(\rho _{A})$ is 
\begin{equation}
\frac{dS(\rho _{A})}{d\alpha _{A}}=-\left[\frac{d\eta _{A}^{(1)}}{d\alpha
_{A}}(1+\ln \eta _{A}^{(1)})+\frac{d\eta _{A}^{(2)}}{d\alpha _{A}}(1+\ln
\eta _{A}^{(2)})\right],  \label{ex-2}
\end{equation}%
where

\begin{equation}
\frac{d\eta _{A}^{(1)}}{d\alpha _{A}}=-\frac{1}{\sqrt{1-4\alpha _{A}}},\frac{%
d\eta _{A}^{(2)}}{d\alpha _{A}}=\frac{1}{\sqrt{1-4\alpha _{A}}}.
\label{ex-3}
\end{equation}%
Thus, 
\begin{eqnarray}
\frac{dS(\rho _{A})}{d\alpha _{A}} &=&-\left[-\frac{1+\ln \eta _{A}^{(1)}}{%
\sqrt{1-4\alpha _{A}}}+\frac{1+\ln \eta _{A}^{(2)}}{\sqrt{1-4\alpha _{A}}}%
\right]  \notag  \label{ex-4} \\
&=&-\frac{1}{\sqrt{1-4\alpha _{A}}}\ln \frac{\eta _{A}^{(2)}}{\eta _{A}^{(1)}%
}.  \label{ex-5-1-}
\end{eqnarray}

Similarly, we obtain 
\begin{eqnarray}
\frac{dS(\rho _{B})}{d\alpha _{B}} &=&-\frac{1}{\sqrt{1-4\alpha _{B}}}\ln 
\frac{\eta _{B}^{(2)}}{\eta _{B}^{(1)}},  \label{ex-5-2} \\
\frac{dS(\rho _{C})}{d\alpha _{C}} &=&-\frac{1}{\sqrt{1-4\alpha _{C}}}\ln 
\frac{\eta _{C}^{(2)}}{\eta _{C}^{(1)}}.  \label{ex-5-3}
\end{eqnarray}

From Eqs. (\ref{der-1}, \ref{ex-5-1-}, \ref{ex-5-2}, \ref{ex-5-3}), we obtain

\begin{eqnarray}
\frac{\partial S(\rho _{A})}{\partial \lambda _{0}} &=&-\frac{2\lambda
_{0}(\lambda _{2}^{2}+\lambda _{3}^{2})}{\sqrt{1-4\alpha _{A}}}\ln \frac{%
\eta _{A}^{(2)}}{\eta _{A}^{(1)}},  \label{ex-6} \\
\frac{\partial S(\rho _{B})}{\partial \lambda _{0}} &=&-\frac{2\lambda
_{0}\lambda _{3}^{2}}{\sqrt{1-4\alpha _{B}}}\ln \frac{\eta _{B}^{(2)}}{\eta
_{B}^{(1)}},  \label{ex-7} \\
\frac{\partial S(\rho _{C})}{\partial \lambda _{0}} &=&-\frac{2\lambda
_{0}\lambda _{2}^{2}}{\sqrt{1-4\alpha _{C}}}\ln \frac{\eta _{C}^{(2)}}{\eta
_{C}^{(1)}}.  \label{ex-8}
\end{eqnarray}

From Eqs. (\ref{ex-1-2}, \ref{ex-6}, \ref{ex-7}, \ref{ex-8}) and $\frac{%
\partial F}{\partial \lambda _{0}}=0$, we obtain 
\begin{eqnarray}
\ell &=&\frac{1}{3}\left(\frac{\lambda _{2}^{2}+\lambda _{3}^{2}}{\sqrt{%
1-4\alpha _{A}}}\ln \frac{\eta _{A}^{(2)}}{\eta _{A}^{(1)}} \right.  \notag
\\
&&+\frac{\lambda _{3}^{2}}{\sqrt{1-4\alpha _{B}}}\ln \frac{\eta _{B}^{(2)}}{%
\eta _{B}^{(1)}}  \notag \\
&& \left. +\frac{\lambda _{2}^{2}}{\sqrt{1-4\alpha _{C}}}\ln \frac{\eta
_{C}^{(2)}}{\eta _{C}^{(1)}} \right).  \label{ex-10}
\end{eqnarray}

Similarly, we consider 
\begin{equation}
\frac{\partial F}{\partial \lambda _{2}}=\frac{1}{3}\left( \frac{\partial
S(\rho _{A})}{\partial \lambda _{2}}+\frac{\partial S(\rho _{B})}{\partial
\lambda _{2}}+\frac{\partial S(\rho _{C})}{\partial \lambda _{2}}\right)
+2\ell \lambda _{2}.  \label{ex-11}
\end{equation}

Clearly,

\begin{equation}
\frac{\partial S(\rho _{\mu })}{\partial \lambda _{2}}=\frac{dS(\rho _{\mu })%
}{d\alpha _{\mu }}\frac{\partial \alpha _{\mu }}{\partial \lambda _{2}},\mu
\in \{A,B,C\}.  \label{ex-14}
\end{equation}

From Eqs. (\ref{ex-5-1-}, \ref{ex-5-2}, \ref{ex-5-3}, \ref{ex-11}, \ref%
{ex-14}) and $\frac{\partial F}{\partial \lambda _{2}}=0$, we obtain

\begin{eqnarray}
\ell &=&\frac{1}{3}\left(\frac{\lambda _{0}^{2}}{\sqrt{1-4\alpha _{A}}}\ln 
\frac{\eta _{A}^{(2)}}{\eta _{A}^{(1)}} \right.  \notag \\
&&+\frac{\lambda _{3}^{2}}{\sqrt{1-4\alpha _{B}}}\ln \frac{\eta _{B}^{(2)}}{%
\eta _{B}^{(1)}}  \notag \\
&& \left. +\frac{\lambda _{0}^{2}+\lambda _{3}^{2}}{\sqrt{1-4\alpha _{C}}}%
\ln \frac{\eta _{C}^{(2)}}{\eta _{C}^{(1)}} \right).  \label{ex-15}
\end{eqnarray}

Similarly, from $\frac{\partial F}{\partial \lambda _{3}}=0$ we obtain 
\begin{eqnarray}
\ell &=&\frac{1}{3}\left(\frac{\lambda _{0}^{2}}{\sqrt{1-4\alpha _{A}}}\ln 
\frac{\eta _{A}^{(2)}}{\eta _{A}^{(1)}} \right.  \notag \\
&&+\frac{\lambda _{0}^{2}+\lambda _{2}^{2}}{\sqrt{1-4\alpha _{B}}}\ln \frac{%
\eta _{B}^{(2)}}{\eta _{B}^{(1)}}  \notag \\
&& \left. +\frac{\lambda _{2}^{2}}{\sqrt{1-4\alpha _{C}}}\ln \frac{\eta
_{C}^{(2)}}{\eta _{C}^{(1)}}\right).  \label{ex-16}
\end{eqnarray}

From Eqs. (\ref{const-1}, \ref{ex-10}, \ref{ex-15}), we obtain 
\begin{equation}
\frac{1-2\lambda _{0}^{2}}{\sqrt{1-4\alpha _{A}}}\ln \frac{\eta _{A}^{(2)}}{%
\eta _{A}^{(1)}}=\frac{1-2\lambda _{2}^{2}}{\sqrt{1-4\alpha _{C}}}\ln \frac{%
\eta _{C}^{(2)}}{\eta _{C}^{(1)}}.  \label{ex-17}
\end{equation}

From Eqs. (\ref{const-1}, \ref{ex-10}, \ref{ex-16}), \ we obtain 
\begin{equation}
\frac{1-2\lambda _{0}^{2}}{\sqrt{1-4\alpha _{A}}}\ln \frac{\eta _{A}^{(2)}}{%
\eta _{A}^{(1)}}=\frac{1-2\lambda _{3}^{2}}{\sqrt{1-4\alpha _{B}}}\ln \frac{%
\eta _{B}^{(2)}}{\eta _{B}^{(1)}}.  \label{ex-18}
\end{equation}

From Eqs. (\ref{const-1}, \ref{ex-15}, \ref{ex-16}), we obtain

\begin{equation}
\frac{1-2\lambda _{3}^{2}}{\sqrt{1-4\alpha _{B}}}\ln \frac{\eta _{B}^{(2)}}{%
\eta _{B}^{(1)}}=\frac{1-2\lambda _{2}^{2}}{\sqrt{1-4\alpha _{C}}}\ln \frac{%
\eta _{C}^{(2)}}{\eta _{C}^{(1)}}.  \label{ex-19}
\end{equation}

When $\lambda _{0}=\lambda _{2}=\lambda _{3}$, Eqs. (\ref{ex-17}, \ref{ex-18}%
, \ref{ex-19}) hold. Via Eq. (\ref{const-1}), one can see that $\lambda
_{0}=\lambda _{2}=\lambda _{3}=1/\sqrt{3}$ is an extrema of $m$ with the
constraint $\sum_{i=0}^{3}\lambda _{i}^{2}=1$.

\section{Appendix D. CKW inequalities for GHZ SLOCC\ class.}

\setcounter{equation}{0} \renewcommand{\theequation}{D\arabic{equation}}

Note that for the GHZ SLOCC\ class, $\lambda _{0}\lambda _{4}\neq 0$. There
is also an additional constraint $\sum_{i=0}^{4}\lambda _{i}^{2}=1$. \vspace{%
10pt}

(A) When $\lambda _{1}=\lambda _{3}=0$ and $\lambda _{2}\neq 0$, we have 
\begin{equation*}
A=\frac{4}{3}(\lambda _{0}^{2}\lambda _{2}^{2}).
\end{equation*}
Clearly, 
\begin{equation*}
A=\frac{4}{3}(\lambda _{0}^{2}\lambda _{2}^{2})\leq \frac{4}{3}\left( \frac{%
\lambda _{0}^{2}+\lambda _{2}^{2}}{2}\right) ^{2}=\frac{1}{3}\left(
1-\lambda _{4}^{2}\right) ^{2}<\frac{1}{3}
\end{equation*}

Similarly, we can obtain $A<\frac{1}{3}$ \ for the case with $\lambda
_{1}=\lambda _{2}=0,\lambda _{3}\neq 0$\ and the case with $\lambda
_{2}=\lambda _{3}=0,\lambda _{1}\neq 0$.

\vspace{10pt} (B) When $\lambda _{1}\lambda _{3}\neq 0$ and $\lambda _{2}=0$%
, $A$ is reduced to 
\begin{equation*}
A=\frac{4}{3}(\lambda _{0}^{2}\lambda _{3}^{2}+\lambda _{1}^{2}\lambda
_{4}^{2}).
\end{equation*}
In light of constrained extreme theorem, we consider the following function 
\begin{equation*}
U=\frac{4}{3}(\lambda _{0}^{2}\lambda _{3}^{2}+\lambda _{1}^{2}\lambda
_{4}^{2})+q(\lambda _{0}^{2}+\lambda _{1}^{2}+\lambda _{3}^{2}+\lambda
_{4}^{2}-1).
\end{equation*}
From $\frac{\partial U}{\partial \lambda _{i}}=0$, obtain only one extreme $%
\lambda _{0}=\lambda _{1}=\lambda _{3}=\lambda _{4}=\frac{1}{2}$, i.e. $%
\frac{1}{2}(|000\rangle +|100\rangle +|110\rangle +|111\rangle )$, at which $%
\max A=\frac{1}{6}$.

\vspace{10pt} (C) When $\lambda _{1}\lambda _{2}\neq 0$ and $\lambda _{3}=0$%
, $A$ is reduced to 
\begin{equation*}
A=\frac{4}{3}(\lambda _{0}^{2}\lambda _{2}^{2}+\lambda _{1}^{2}\lambda
_{4}^{2}).
\end{equation*}
The discussion is similar to (B), there is only one extreme $\lambda
_{0}=\lambda _{1}=\lambda _{2}=\lambda _{4}=\frac{1}{2}$, i.e. $\frac{1}{2}%
(|000\rangle +|100\rangle +|101\rangle +|111\rangle )$, $\max A=\frac{1}{6}$.

\vspace{10pt}

(D) When $\lambda _{1}=0$ and $\lambda _{2}\lambda _{3}\neq 0$, then 
\begin{equation*}
A=\frac{4}{3}(\lambda _{0}^{2}\lambda _{3}^{2}+\lambda _{0}^{2}\lambda
_{2}^{2}+\lambda _{2}^{2}\lambda _{3}^{2}).
\end{equation*}

The constraint is 
\begin{equation*}
\lambda _{0}^{2}+\lambda _{2}^{2}+\lambda _{3}^{2}+\lambda _{4}^{2}=1,
\end{equation*}%
where $\lambda _{4}$ is considered as a parameter. In light of constrained
extreme theorem, for a fixed $\lambda _{4}$,\ when $\lambda _{0}=\lambda
_{2}=\lambda _{3}$, i.e., $\lambda _{0}|000\rangle +\lambda _{0}|101\rangle
+\lambda _{0}|110\rangle +\lambda _{4}|111\rangle $, then $\ A$ has the
maximum $A=4\lambda _{0}^{4}=\frac{4}{9}(1-\lambda _{4}^{2})^{2}<\frac{4}{9}$%
.

\vspace{10pt} (E) When $\lambda _{1}\lambda _{2}\lambda _{3}\neq 0$, let 
\begin{equation*}
f=\frac{4}{3}(\lambda _{0}^{2}\lambda _{3}^{2}+\lambda _{0}^{2}\lambda
_{2}^{2}+(\lambda _{1}\lambda _{4}+\lambda _{2}\lambda _{3})^{2}).
\end{equation*}

Clearly, $A\leq f$. We next calculate the maximum value of $f$. The
constraint reads $\lambda _{0}^{2}+\lambda _{1}^{2}+\lambda _{2}^{2}+\lambda
_{3}^{2}+\lambda _{4}^{2}=1.$ In light of the constrained extreme theorem,
we consider the function 
\begin{equation*}
F=f+g(\lambda _{0}^{2}+\lambda _{1}^{2}+\lambda _{2}^{2}+\lambda
_{3}^{2}+\lambda _{4}^{2}-1).
\end{equation*}

From $\frac{\partial F}{\partial \lambda _{i}}=0$, $i=0,1,2,3,4,$ we obtain $%
\lambda _{1}=\lambda _{2}=\lambda _{3}=\lambda _{4}$ and $\lambda _{0}=0$.
Thus, $f$ has the maximum value $\frac{1}{3}$. However, we require that $%
\lambda _{0}$ do not vanish. From $\sum_{i=0}^{4}\lambda _{i}^{2}=1$, we get 
$\lambda _{0}^{2}+4\lambda _{4}^{2}=1$. Thus, when $\lambda _{1}=\lambda
_{2}=\lambda _{3}=\lambda _{4}$, $A=\frac{4}{3}((2\lambda
_{4}^{2}))(1-2\lambda _{4}^{2})<1/3.$ Clearly, $\lim_{\lambda
_{0}\rightarrow 0}A=\frac{1}{3}$. For example, when $\lambda _{0}^{2}=\frac{1%
}{100}$, then $A=\allowbreak \frac{3333}{10\,000}\approx \frac{1}{3}$.

\textbf{Data Availability Statement:} The datasets generated during and/or
analyzed during the current study are available from the corresponding
author on reasonable request.

\textbf{Conflicts of interest/Competing interests:} The authors have no
conflicts of interest to declare that are relevant to the content of this
article.

\end{document}